# Exciton-Phonon Dynamics with Long-Range Interaction

**Nick Laskin**

TopQuark Inc., Toronto, Canada
(e-mail: nlaskin@rocketmail.com)

**Abstract** Exciton-phonon dynamics on a 1D lattice with long-range exciton-exciton interaction have been introduced and elaborated. Long-range interaction leads to a nonlocal integral term in the motion equation of the exciton subsystem if we go from discrete to continuous space. In some particular cases for power-law interaction, the integral term can be expressed through a fractional order spatial derivative. A system of two coupled equations has been obtained, one is a fractional differential equation for the exciton subsystem, the other is a standard differential equation for the phonon subsystem. These two equations present a new fundamental framework to study nonlinear dynamics with long-range interaction. New approaches to model the impact of long-range interaction on nonlinear dynamics are: fractional generalization of Zakharov system, Hilbert-Zakharov system, Hilbert-Ginzburg-Landau equation and nonlinear Hilbert-Schrödinger equation. Nonlinear fractional Schrödinger equation and fractional Ginzburg-Landau equation are also part of this framework.

## 1.1 INTRODUCTION

Dynamic lattice models are widely used to study a broad set of physical phenomena and systems. In the early 1970's a novel mechanism for the localization and transport of vibrational energy in certain types of molecular chains was proposed by A.S. Davydov (Davydov 1973). He pioneered the concept of *solitary excitons* or the *Davydov soliton* (Davydov and Kislukha 1973). His theoretical model to study solitary excitons is based on exciton-phonon lattice dynamics with nearest-neighbour exciton-exciton interaction, the so-called *Davydov model*. Our primary focus is the analytical developments of quantum 1D exciton-phonon dynamics with power-law long-range exciton-exciton interaction $J_{n,m} = J/|n-m|^{-s}$, $(s > 0)$ for excitons located at lattice sites *n* and *m*. In addition to the well-known interactions with integer values of *s*, some complex media can be described by fractional values of *s* (see, for example, references in (Zaslavsky et al. 2006)). Using the ideas first developed in (Laskin and Zaslavsky 2006), we elaborate the Davydov model for an exciton-phonon system with a fractional power-law exciton-exciton interaction. It has been shown that 1D lattice exciton-phonon dynamics in the long-wave limit can be effectively presented by the general system of two coupled equations for exciton and phonon dynamic variables. The dynamic equation describing the exciton subsystem is the fractional differential equation, which is a manifestation of non-locality of interaction, originating from the long-range interaction term. These two dynamic equations can be considered as a new general approach to study nonlinear dynamics with long-range interaction. Particular physical cases include: nonlinear fractional Schrödinger equation, fractional Ginzburg-Landau equation, fractional generalization of Zakharov system, Hilbert-Zakharov system, Hilbert-Ginzburg-Landau equation and nonlinear Hilbert-Schrödinger equation.

The paper is organized as follows. In Sec.2 we generalize Davydov's Hamiltonian for the case of long-range power-law exciton-exciton interaction. The system of two coupled discrete equations of motion for exciton and phonon subsystems has been found using the Davydov anzatz. Transformation to the system of two continuous equations of motion has been performed in the long wave limit. Sec.3 focuses on new nonlinear fractional differential equations resulting from our general approach to study the 1D exciton-phonon system with long-range interaction. In conclusion, we outline our new developments.

## 1.2 LATTICE EXCITON-PHONON HAMILTONIAN WITH LONG-RANGE INTERACTION

### 1.2.1 Davydov's Hamiltonian

To model 1D quantum lattice dynamics with long-range exciton-exciton interaction we follow (Davydov 1991) and consider a linear, rigid arrangement of sites with one molecule at each lattice site. Davydov's Hamiltonian reads

$$H = H_{ex} + H_{ph} + H_{int}. \qquad (1)$$

Here $H_{ex}$ is the Hamiltonian operator of the exciton system, which describes dynamics of intra-molecular excitations or simply excitons, $H_{ph}$ is phonon Hamiltonian operator, which describes molecular displacements or, in other words, the lattice vibrations, and $H_{int}$ is the exciton-phonon operator, which describes the interaction of an exciton with lattice vibrations. The exciton Hamiltonian is

$$H_{ex} = \varepsilon \sum_{n=-\infty}^{\infty} b_n^+ b_n - \sum_{n,m=-\infty}^{\infty} J_{n,m} b_n^+ b_m, \qquad (2)$$

where $b_n^+$ is creation and $b_n$ is annihilation operators of an exciton on the $n$ site. Operators $b_n^+$ and $b_n$ satisfy the relations $[b_n, b_m^+] = \delta_{n,m}$, $[b_n, b_m] = 0$, $[b_n^+, b_m^+] = 0$.

Parameter $\varepsilon$ is exciton energy on the site, $J_{n,m}$ is the exciton transfer matrix, which describes exciton-exciton interaction between sites $n$ and $m$. To extend

Davydov's model and go beyond the nearest-neighbour interaction, we introduce the power-law interaction between excitons on sites *n* and *m*

$$J_{n,m} = J_{n-m} = \frac{J}{|n-m|^s}, \quad n \neq m, \qquad (3)$$

where *J* is the interaction constant, parameter *s* covers different physical models; the nearest-neighbour approximation ($s=\infty$), the dipole-dipole interaction ($s=3$), and the Coulomb potential ($s=1$). Our main interest will be in fractional values of *s* that can appear for more sophisticated interaction potentials attributed to complex media.

The phonon Hamiltonian $H_{ph}$ is

$$H_{ph} = \sum_{n=-\infty}^{\infty} \left( \frac{\hat{p}_n^2}{2m} + \frac{w}{2} (\hat{u}_{n+1} - \hat{u}_n)^2 \right), \qquad (4)$$

where *w* is the elasticity constant of the 1D lattice, $\hat{u}_n$ is the displacement operator from the equilibrium position of site *n*, $\hat{p}_n$ is the momentum operator of site *n*, and *m* is molecular mass.

Finally, the exciton-phonon Hamiltonian $H_{int}$ is

$$H_{int} = \chi \sum_{n=-\infty}^{\infty} (\hat{u}_{n+1} - \hat{u}_n) b_n^+ b_n, \qquad (5)$$

with coupling constant $\chi$. Furthermore, aiming to obtain a system of dynamic equations for the exciton-photon system under consideration, we introduce Davydov's ansatz.

1.2.2 Davydov's anzatz and motion equations

To study system (1) we introduce quantum state vector $|\phi(t)>$ following (Davydov 1973, 1991), (Scott 1992)

$$|\phi(t)> = |\Psi(t)>|\Phi(t)>, \qquad (6)$$

where quantum vectors $|\Psi(t)>$ and $|\Phi(t)>$ are defined by

$$|\Psi(t)> = \sum_n \psi_n(t) b_n^+ |0>_{ex}, \qquad (7)$$

and

$$|\Phi(t)> = \exp\left\{ -\frac{i}{\hbar} \sum_n (\xi_n(t) \hat{p}_n - \eta_n(t) \hat{u}_n) \right\} |0>_{ph}, \qquad (8)$$

here, $\hbar$ is the Planck's constant, $|0>_{ex}$ and $|0>_{ph}$ are vacuum states of exciton and phonon subsystems, and $\xi_n(t)$ is the diagonal matrix element of the displacement operator $\hat{u}_n$ in the basis defined by (6), while $\eta_n(t)$ is diagonal matrix element of the momentum operator $\hat{p}_n$ in the same basis,

$$\xi_n(t) = <\phi(t)|\hat{u}_n|\phi(t)>, \quad \eta_n(t) = <\phi(t)|\hat{p}_n|\phi(t)>.$$

State vector $|\phi(t)>$ satisfies the normalization condition

$$<\phi(t)|\phi(t)> = \sum_n |\psi_n(t)|^2 = N,$$

with $|\psi_n(t)|^2$ being the probability to find exciton on the $n^{\text{th}}$ site and $N$ is the total number of excitons.

Therefore, the study of dynamics of an exciton-photon system (1) can be performed in terms of the functions $\psi_n(t)$, $\xi_n(t)$ and $\eta_n(t)$. In other words, *Davydov's ansatz* defined by (6)-(8) allows us to go from the quantum Hamiltonian operator introduced by (1) to the Hamiltonian function developed below. In the basis of vectors $|\phi(t)>$, Hamiltonians $H_{ex}$, $H_{ph}$, and $H_{int}$ become functions of dynamic variables $\psi_n(t)$, $\psi_n^*(t)$, $\xi_n(t)$ and $\eta_n(t)$

$$<\phi(t)|H_{ex}|\phi(t)> = H_{ex}(\psi_n, \psi_n^*) =$$
$$\varepsilon \sum_{n=-\infty}^{\infty} \psi_n^*(t)\psi_n(t) - \sum_{n,m=-\infty}^{\infty} J_{n-m}\psi_n^*(t)\psi_m(t), \tag{9}$$

and

$$<\phi(t)|H_{ph}|\phi(t)> = H_{ph}(\xi_n, \eta_n) =$$
$$\sum_{n=-\infty}^{\infty} \left( \frac{\eta_n^2}{2m} + \frac{w}{2}(\xi_{n+1} - \xi_n)^2 \right), \tag{10}$$

and

$$<\phi(t)|H_{int}|\phi(t)> = H_{int}(\psi_n, \psi_n^*; \xi_n, \eta_n) =$$
$$\chi \sum_{n=-\infty}^{\infty} (\xi_{n+1} - \xi_n)\psi_n^*(t)\psi_n(t), \tag{11}$$

From (9)-(11) we obtain the system of dynamic equations in discrete space for $\psi_n(t)$, $\xi_n(t)$ and $\eta_n(t)$,

$$i\hbar \frac{\partial \psi_n(t)}{\partial t} = \Lambda \psi_n(t) -$$

$$-\sum_{\substack{m \\ (n \neq m)}} J_{n-m} \psi_m(t) + \chi(\xi_{n+1}(t) - \xi_n(t))\psi_n(t), \qquad (12)$$

$$\frac{\partial \xi_n(t)}{\partial t} = \frac{\eta_n(t)}{m}, \qquad (13)$$

and

$$\frac{\partial \eta_n(t)}{\partial t} = w(\xi_{n+1}(t) - 2\xi_n(t) + \xi_{n-1}(t)) +$$
$$+ \chi(|\psi_{n+1}(t)|^2 - |\psi_n(t)|^2), \qquad (14)$$

where constant $\Lambda$ is

$$\Lambda = \varepsilon + \sum_{n=-\infty}^{\infty} \left( \frac{m}{2}\left(\frac{\partial \xi_n(t)}{\partial t}\right)^2 + \frac{w}{2}(\xi_{n+1}(t) - \xi_n(t))^2 \right).$$

Substituting $\eta_n(t)$ from (13) into (14) yields

$$m\frac{\partial^2 \xi_n(t)}{\partial t^2} = w(\xi_{n+1}(t) - 2\xi_n(t) + \xi_{n-1}(t)) +$$
$$+ \chi(|\psi_{n+1}(t)|^2 - |\psi_n(t)|^2). \qquad (15)$$

Our focus now is the system of two coupled discrete dynamic equations (12) and (15).

1.2.3 From lattice to continuum

To go from the discrete to continuum version of (12) and (15) let us introduce

$$\varphi(k,t) = \sum_{n=-\infty}^{\infty} e^{-ikn} \psi_n(t), \qquad v(k,t) = \sum_{n=-\infty}^{\infty} e^{-ikn} \xi_n(t),$$

where $\psi_n(t)$ is related to $\varphi(k,t)$ as

$$\psi_n(t) = \frac{1}{2\pi} \int_{-\pi}^{\pi} dk\, e^{ikn} \varphi(k,t),$$

and $\xi_n(t)$ is related to $v(k,t)$ as

$$\xi_n(t) = \frac{1}{2\pi} \int_{-\pi}^{\pi} dk\, e^{ikn} v(k,t),$$

and $k$ can be considered as a wave number. In the long wave limit when the wavelength exceeds the intersite scale $a$ (let's put for simplicity $a=1$) we may consider $\varphi(k,t)$ as a $k^{\underline{th}}$ Fourier component of continuous function $\psi(x,t)$, $\psi_n(t)\xrightarrow[k\to 0]{}\psi(x,t)$ and $v(k,t)$ as a $k^{\underline{th}}$ Fourier component of function $\xi(x,t)$, $\xi_n(t)\xrightarrow[k\to 0]{}\xi(x,t)$. That is, functions $\psi(x,t)$ and $\varphi(k,t)$ are related to each other by the Fourier transforms

$$\psi(x,t) = \frac{1}{2\pi}\int_{-\infty}^{\infty} dk\, e^{ikx}\varphi(k,t), \quad \varphi(k,t) = \int_{-\infty}^{\infty} dx\, e^{ikx}\psi(x,t),$$

and similarly for $\xi(x,t)$ and $v(k,t)$,

$$\xi(x,t) = \frac{1}{2\pi}\int_{-\infty}^{\infty} dk\, e^{ikx} v(k,t), \quad v(k,t) = \int_{-\infty}^{\infty} dx\, e^{ikx}\xi(x,t).$$

Therefore, we conclude that in the long wave limit (12) and (15) become continuous equations of motion

$$i\hbar\frac{\partial \psi(x,t)}{\partial t} = \lambda\psi(x,t) -$$
$$- \int_{-\infty}^{\infty} dy\, \partial_x K(x-y)\partial_x \psi(x,t) + \chi\frac{\partial \xi(x,t)}{\partial x}\psi(x,t), \qquad (16)$$

and

$$m\frac{\partial^2 \xi(x,t)}{\partial t^2} = w\frac{\partial^2 \xi(x,t)}{\partial x^2} + 2\chi\frac{\partial |\psi(x,t)|^2}{\partial x}, \qquad (17)$$

where kernel $K(x)$ in (16) has been introduced as

$$K(x) = \frac{1}{\pi}\int_{-\infty}^{\infty} dk\, e^{ikx}\frac{G(k)}{k^2},$$

with function $G(k)$ defined by

$$G(k) = J(0) - J(k), \qquad J(k) = \sum_{\substack{n=-\infty \\ n\neq 0}}^{\infty} e^{-ikn} J_n,$$

here, $J_n$ is given by (3), and finally, $\lambda = \Lambda - J(0)$.

Thus, we obtained a new system of coupled dynamic equations (16) and (17) to model 1D exciton-phonon dynamics with long-range exciton-exciton interaction (3). Field $\psi(x,t)$ describes the exciton subsystem and field $\xi(x,t)$ describes the phonon subsystem. Equation (16) is the integro-differential equation while equation (17) is the differential one. The integral term in (16), which is a manifestation of non-locality of interaction, comes from the long-range interaction term in Hamiltonian (2).

**1.3 FRACTIONAL DIFFERENTIAL EQUATIONS TO STUDY EXCITON-PHONON DYNAMICS**

To transform system (16), (17) into the system of coupled differential equations we use the properties of function $G(k)$ at limit $k \to 0$, which can be obtained from the polylogarithm asymptotics (Laskin and Zaslavsky 2006)

$$G(k) \sim D_s \, |k|^{s-1}, \quad 2 \leq s < 3, \tag{18}$$

$$G(k) \sim -Jk^2 \ln k, \quad s = 3, \tag{19}$$

$$G(k) \sim \frac{J\zeta(s-2)}{2} k^2, \quad s > 3, \tag{20}$$

where $\Gamma(s)$ is $\Gamma$-function, $\zeta(s)$ is the Riemann zeta function and coefficient $D_s$ is defined by

$$D_s = \frac{\pi J}{\Gamma(s)\sin(\pi(s-1)/2)}. \tag{21}$$

It is seen from (18) that the fractional power of $k$ occurs for interactions with $2 < s < 3$ only. In the coordinate space, fractional power of $|k|$ gives us the fractional Riesz derivative (Samko et al. 1993), (Saichev and Zaslavsky 1997), and we come to a fractional differential equation

$$i\hbar \frac{\partial \psi(x,t)}{\partial t} = \lambda \psi(x,t) -$$

$$- D_s \partial_x^{s-1} \psi(x,t) + \chi \frac{\partial \xi(x,t)}{\partial x} \psi(x,t), \quad 2 < s < 3, \tag{22}$$

where, $\partial_x^{s-1}$ is the Riesz fractional derivative of order $s-1$

$$\partial_x^{s-1} \psi(x,t) = -\frac{1}{2\pi} \int_{-\infty}^{\infty} dk\, e^{ikx} \, |k|^{s-1} \, \varphi(k,t).$$

Thus, our main result is the new system of coupled equations (17) and (22) to study exciton-phonon dynamics with long-range interaction on a 1D lattice. The system of equations (17) and (22) is in fact a new general framework to model nonlinear dynamic phenomena with long-range interaction.

Now let us introduce and briefly discuss new theoretical approaches originating from the framework. They are: fractional generalization of the Zakharov system, the nonlinear fractional Schrödinger equation, the fractional Ginzburg-Landau equation, the Hilbert-Zakharov system, the nonlinear Hilbert-Schrödinger equation, and the fractional Hilbert-Ginzburg-Landau equation.

1.3.1 Fractional generalization of Zakharov system

Introducing a new variable $\sigma(x,t) = \dfrac{\partial \xi(x,t)}{\partial x}$ turns (22) and (17) into the following new system of equations for fields $\psi(x,t)$ and $\sigma(x,t)$,

$$i\hbar \frac{\partial \psi(x,t)}{\partial t} = \lambda \psi(x,t) -$$

$$- D_s \partial_x^{s-1} \psi(x,t) + \chi \sigma(x,t)\psi(x,t), \quad 2 < s < 3, \tag{23}$$

and

$$\left( \frac{\partial^2}{\partial t^2} - v^2 \frac{\partial^2}{\partial x^2} \right)\sigma(x,t) = \frac{2\chi}{m} \frac{\partial^2}{\partial x^2} |\psi(x,t)|^2, \tag{24}$$

where $v = \sqrt{w/m}$ is the velocity of sound.

If we introduce the $\phi(x,t)$,

$$\phi(x,t) = \exp\{i\lambda t/\hbar\}\psi(x,t), \tag{25}$$

then the system of two coupled equations (23) and (24) reads

$$i\hbar \frac{\partial \phi(x,t)}{\partial t} = -D_s \partial_x^{s-1} \phi(x,t) + \chi \sigma(x,t)\phi(x,t), \quad 2 < s < 3, \tag{26}$$

and

$$\left( \frac{\partial^2}{\partial t^2} - v^2 \frac{\partial^2}{\partial x^2} \right)\sigma(x,t) = \frac{2\chi}{m} \frac{\partial^2}{\partial x^2} |\phi(x,t)|^2, \tag{27}$$

Equations (26) and (27) can be considered as a fractional generalization of the Zakharov system introduced in 1972 to study the Langmuir waves propagation in an ionized plasma (Zakharov 1972).

1.3.2 Nonlinear fractional Schrödinger equation

Assuming the existence of a stationary solution $\partial \xi(x,t)/\partial t=0$ in the system of (17) and (22) results in the following fractional differential equation for wave function $\psi(x,t)$,

$$i\hbar \frac{\partial \psi(x,t)}{\partial t} = \lambda \psi(x,t) -$$

9$$-D_s\partial_x^{s-1}\psi(x,t)-\frac{2\chi}{w}|\psi(x,t)|^2\psi(x,t), \quad 2<s<3, \qquad (28)$$

which can be rewritten in the form of a nonlinear fractional Schrödinger equation,

$$i\hbar\frac{\partial\phi(x,t)}{\partial t}=-D_s\partial_x^{s-1}\phi(x,t)-\frac{2\chi}{w}|\phi(x,t)|^2\phi(x,t), \qquad (29)$$

where $2<s<3$ and the wave function $\phi(x,t)$ is related to the wave function $\psi(x,t)$ by means of (25).

It follows from (20) that for $s>3$, (29) turns into the nonlinear Schrödinger equation

$$i\hbar\frac{\partial\phi(x,t)}{\partial t}=-\frac{J\zeta(s-2)}{2}\partial_x^2\phi(x,t)-\frac{2\chi}{w}|\phi(x,t)|^2\phi(x,t).$$

where $\partial_x^2=\partial^2/\partial x^2$.

Finally, note that the linear fractional Schrödinger equation in one and three dimensions has been developed at first in (Laskin 2000a, b, c, 2002). Three quantum mechanical problems were studied in these papers; a quantum particle in an infinite potential well, fractional quantum oscillator, and fractional Bohr atom. The energy spectra for these three fractional quantum mechanical problems were found using the linear fractional Schrödinger equation.

1.3.3 Fractional Ginzburg-Landau equation

In the case of propagating waves we can search for the solution of system (17) and (22) in the form of travelling waves, $\psi(x,t) = \psi(x\text{-v}t)$, and $\xi(x,t) = \xi(x\text{-v}t)$, where v is velocity of the wave. From (17) and (22) let's go to (23) and (24) and substitute $\psi(x,t)=\psi(\zeta)$, and $\sigma(x,t)=\sigma(\zeta)$, where $\zeta=x\text{-v}t$. It is easy to see that the solution of (24) is

$$\sigma(x,t)=\frac{2\chi}{m(\text{v}^2-v^2)}|\psi(\zeta)|^2. \qquad (30)$$

Then (23) results in nonlinear equation

$$i\hbar\text{v}\frac{\partial\psi(\zeta)}{\partial\zeta}=\lambda\psi(\zeta)-$$

$$-D_s\partial_\zeta^{s-1}\psi(\zeta)+\gamma|\psi(\zeta)|^2\psi(\zeta), \quad 2<s<3, \qquad (31)$$

where $\gamma$ is the nonlinearity parameter

$$\gamma=\frac{2\chi}{m(\text{v}^2-v^2)}. \qquad (32)$$

Introducing wave function $\phi(\zeta)$ related to wave function $\psi(\zeta)$ by

$$\phi(\zeta) = \exp\{i\lambda\zeta / \hbar \mathrm{v}\}\psi(\zeta),$$

results in fractional Ginzburg-Landau equation

$$i\hbar \mathrm{v} \frac{\partial \phi(\zeta)}{\partial \zeta} = -D_s \partial_\zeta^{s-1} \phi(\zeta) + \gamma |\phi(\zeta)|^2 \phi(\zeta), \quad 2 < s < 3,$$

which was initially proposed in (Weitzner and Zaslavsky 2003).

1.3.4 Hilbert-Zakharov system

It follows from (18) that in the case when $s=2$, the function $G(k)$ at limit $k \to 0$ takes the form

$$G(k) \sim \pi J |k|, \quad s = 2.$$

Hence, (23) becomes

$$i\hbar \frac{\partial \psi(x,t)}{\partial t} = \lambda \psi(x,t) -$$

$$- \pi J \mathcal{H}\{\partial_x \psi(x,t)\} + \chi \sigma(x,t) \psi(x,t), \quad s = 2, \qquad (33)$$

here $\mathcal{H}$ is the Hilbert integral transform defined by

$$\mathcal{H}\{\varphi(x,t)\} = P \int_{-\infty}^{\infty} dy \frac{\varphi(y,t)}{y - x},$$

where $P$ stands for the Cauchy principal value of the integral.

We will call the system of equations (33) and (24) as the Hilbert-Zakharov system.

1.3.5 Nonlinear Hilbert-Schrödinger equation

In the case when $s=2$ and $\partial \xi(x,t)/\partial t = 0$, the system of equations (16) and (17) results in the following nonlinear quantum mechanical equation for wave function $\psi(x,t)$,

$$i\hbar \frac{\partial \psi(x,t)}{\partial t} = \lambda \psi(x,t) -$$

$$- \pi J \mathcal{H}\{\partial_x \psi(x,t)\} + \frac{2\chi}{m} |\psi(x,t)|^2 \psi(x,t). \qquad (34)$$

If we introduce wave function $\phi(x,t)$ related to wave function $\psi(x,t)$ by means of (25), then we have nonlinear Hilbert-Schrödinger equation

$$i\hbar \frac{\partial \phi(x,t)}{\partial t} = -\pi J \mathcal{H}\{\partial_x \phi(x,t)\} - \frac{2\chi}{\omega}|\phi(x,t)|^2 \phi(x,t). \quad (35)$$

This equation was first developed in (Laskin and Zaslavsky 2006).

1.3.6 Hilbert-Ginzburg-Landau equation

In the case when $s$=2, let's search for the solution of the system of equations (33) and (24) in the form of travelling waves, $\psi(x,t)= \psi(x\text{-}vt)$, and $\xi(x,t)= \xi(x\text{-}vt)$, where v is velocity of the wave. In this case the solution of (24) has the form of (30). Thus, (31) results in

$$i\hbar v \frac{\partial \psi(\zeta)}{\partial \zeta} = \lambda \psi(\zeta) - \pi J \mathcal{H}\{\partial_x \psi(\zeta)\} + \gamma |\psi(\zeta)|^2 \psi(\zeta), \quad (36)$$

where $\gamma$ is the nonlinearity parameter introduced by (32) and $\zeta$=$x$-v$t$.

If we introduce wave function $\phi(x,t)$ related to wave function $\psi(x,t)$ by means of (25), then from (36) we obtain

$$i\hbar v \frac{\partial \phi(\zeta)}{\partial \zeta} = -\pi J \mathcal{H}\{\partial_x \phi(\zeta)\} + \gamma |\phi(\zeta)|^2 \phi(\zeta), \quad (37)$$

We will call (37) the Hilbert-Ginzburg-Landau equation.

**1.4 CONCLUSION**

We have introduced and developed exciton-phonon dynamics with long-range exciton-exciton interaction. It has been shown that the long-range power-law interaction leads, in general, to a nonlocal integral term in the motion equation of an exciton subsystem if we go from a 1D lattice to continuous space. In some particular cases for power-law interaction with non-integer power $s$, the nonlocal integral term can be expressed through a spatial derivative of fractional order. We have obtained the system of two coupled equations, where one is fractional differential equation for exciton subsystem and the other one is a standard differential equation for phonon subsystem. It has been found that this system of two coupled equations can be further simplified to present new fundamental approaches for studying a wide range of nonlinear physical phenomena with long-range interaction. These approaches are: nonlinear fractional Schrödinger equation, fractional Ginzburg-Landau equation, fractional generalization of Zakharov system, Hilbert-Zakharov system, Hilbert-Ginzburg-Landau equation and nonlinear Hilbert-Schrödinger equation.

The results presented will further advance applications of fractional calculus to study waves and chaos phenomena in media with long-range interaction. Imple-mentation of new theoretical approaches will initiate developing numerical algorithms to simulate impact of long-range interaction on nonlinear dynamics.